\def\year{2018}\relax
\documentclass[letterpaper]{article} 
\usepackage{aaai18}  
\usepackage{times}  
\usepackage{helvet}  
\usepackage{courier}  
\usepackage{url}  
\usepackage{graphicx}  
\usepackage{MyDefinitions}
\begin{document}
%
\title{B\"uchi automata augmented with spatial constraints: simulating an alternating with a nondeterministic and deciding the emptiness problem for the latter}
\author{Amar Isli\\
        University of Sciences and Technology Houari Boumedi\`ene\\
	Department of Computer Science\\
	BP 32, DZ-16111 Bab Ezzouar, Algiers\\
	Algeria\\
	a\_isli@yahoo.com\\
}
\maketitle
\begin{abstract}
\footnote{Exactly as rejected by the KR'2018 Conference. The paper, together with another, also rejected by the KR'2018 Conference, had been extracted
from a substantial revision of \cite{Isli03a}. Further revisions are needed before replacing \cite{Isli03a}.}\\
The aim of this work is to thoroughly investigate B\"uchi automata augmented with spatial constraints.
The input trees of such an automaton are infinite $k$-ary $\Sigma$-trees, with the nodes standing for
time points, and $\Sigma$ including, additionally to its uses in classical $k$-ary $\Sigma$-trees, the
description of the snapshot of an $n$-object spatial scene of interest. The constraints, from an RCC8-like
spatial Relation Algebra (RA) x, are used to impose spatial constraints on objects of the spatial scene,
eventually at different nodes of the input trees. We show that a B\"uchi alternating automaton augmented
with spatial constraints can be simulated with a classical B\"uchi nondeterministic automaton of the same
type, augmented with spatial constraints. We then provide a nondeterministic doubly depth-first polynomial
space algorithm for the emptiness problem of the latter automaton. Our main motivation came from another
work, also submitted to this conference, which defines a spatio-temporalisation of the well-known
family $\alcd$ of description logics with a concrete domain: together, the two works provide an effective
solution to the satisfiability problem of a concept of the spatio-temporalisation with respect to a weakly cyclic TBox.

{\bf Author keywords:}
B\"uchi automata,
Alternating automata,
Nondeterministic automata,
Qualitative spatial constraints,
Emptiness problem,
Doubly depth-first search,
Polynomial space algorithm.
\end{abstract}
\newtheorem{theorem}{Theorem}
\newtheorem{conjecture}{Conjecture}
\newtheorem{corollary}{Corollary}
\newtheorem{proposition}{Proposition}
\newtheorem{lemma}{Lemma}
\newtheorem{discussion}{Discussion}
\newtheorem{definition}{Definition}
\newtheorem{remark}{Remark}
\newtheorem{example}{Example}
\section{Introduction}\label{sect1}
The aim of this work is to thoroughly investigate B\"uchi automata augmented with spatial constraints.

The first result we will show is that a B\"uchi alternating automaton augmented
with spatial constraints can be simulated with a classical B\"uchi nondeterministic automaton of the same
type, augmented with spatial constraints.
An algorithm is known from \cite{Isli93a,Isli96a} for the simulation of a B\"uch alternating automaton on infinite words with a B\"uchi nondeterministic automaton. We
adapt it to the simulation of a B\"uchi alternating automaton on $k$-ary $\Sigma$-trees augmented with constraints. The interesting point in this first result is that the adding of spatial constraints does not compromise the simulation method.

The most interesting part of the work is the second result, to the best of our knowledge original, which provides a nondeterministic doubly depth-first polynomial
space algorithm for the emptiness problem of a B\"uchi nondeterministic automaton augmented with spatial constraints. The algorithm is expected to have 
very positive repercussions not only in the Description Logics community, but in many other fields of computer science as well.

Our main motivation came from another
work, also submitted to this conference, which defines a spatio-temporalisation of the well-known
family $\alcd$ of description logics with a concrete domain \cite{BaaderH91a}: together, the two works provide an effective
solution to the satisfiability problem of a concept of the spatio-temporalisation with respect to a weakly cyclic TBox.
\section{Concrete domain}
\begin{definition}[concrete domain \cite{BaaderH91a}]\label{cddefinition}
A concrete domain $\textsl{D}$ consists of a pair
$(\Delta _{\textsl{D}},\Phi _{\textsl{D}})$, where
$\Delta _{\textsl{D}}$ is a set of (concrete) objects, and
$\Phi _{\textsl{D}}$ is a set of predicates over the objects in $\Delta _{\textsl{D}}$.
Each predicate $P\in\Phi _{\textsl{D}}$ is associated
with an arity $n$ and we have
$P\subseteq (\Delta _{\textsl{D}})^n$.
\end{definition}
\begin{definition}[admissibility \cite{BaaderH91a}]\label{cdadmissibility}
A concrete domain $\textsl{D}$ is admissible if:
(1) the set of its predicates is closed under negation and
    contains a predicate for $\Delta _{\textsl{D}}$; and
(2) the satisfiability problem for finite conjunctions of predicates is decidable.
\end{definition}

Any spatial RA $x$ for
    which the atoms are Jointly Exhaustive and Pairwise Disjoint (henceforth JEPD), and
    such that the atomic relations form a decidable subclass, can be used to generate a
concrete domain $\textsl{D}_x$ for members of the family $\xdl$ of qualitative theories
for spatial change. Such a concrete domain is used for representing knowledge on
$p$-tuples of objects of the spatial domain at hand, $p$ being the arity of the $x$ relations;
stated otherwise, the $x$ relations will be used as the predicates of $\textsl{D}_x$.

Let $x\in\{\rcc8 ,\cdalg ,\atra\}$. The concrete domain generated by $x$, $\textsl{D}_x$, can be written as
$\textsl{D}_x=(\Delta _{\textsl{D}_x},\Phi _{\textsl{D}_x})$, with:
                         $\textsl{D}_{\rcc8}   =(\rtopspace ,2^{\rccats})$,
                         $\textsl{D}_{\cdalg}  =(\deuxdp ,2^{\cdalgats})$ and
                         $\textsl{D}_{\atra}  =(\deuxdo ,2^{\atraats})$,
where
$\rtopspace$ is the set of regions of a topological space $\topspace$;
    $\deuxdp$ is the set of 2D points;
    $\deuxdo$ is the set of 2D orientations; and
$\xat$, as we have seen, is the set of $x$ atoms
    ---$2^{\xat}$ is thus the set of all $x$ relations.
%

Admissibility of the concrete domains $\textsl{D}_x$
is an immediate consequence of (decidability and) tractability of the subset
$\{\{r\}|r\in\xat\}$ of $x$ atomic relations, for each
$x\in\{\rcc8 ,\cdalg ,\atra\}$. The reader is referred to \cite{RenzN99a} for $x=\rcc8$, to \cite{Ligozat98a}
for $x=\cdalg$, and to \cite{IsliC98a,IsliC00a} for $x=\atra$.
\section{Alternating automata}
\begin{definition}[free distributive lattice]
Let $S$ be a set of generators. $\textsl{L}(S)$ denotes the free distributive lattice generated by $S$.
$\textsl{L}(S)$ can be thought of as the set of logical formulas built from variables taken from
$S$ using the disjunction and conjunction operators $\vee$ and $\wedge$ (but not the negation
operator $\neg$). In other words, $\textsl{L}(S)$ is the smallest set such that:
\begin{enumerate}
  \item for all $s\in S$, $s\in \textsl{L}(S)$; and
  \item if $e_1$ and $e_2$ belong to $\textsl{L}(S)$, then so do $e_1\wedge e_2$ and
    $e_1\vee e_2$.
\end{enumerate}
\end{definition}

\begin{definition}[set representation]\label{setrepresentation}
Each element $e\in \textsl{L}(S)$ has, up to isomorphism, a unique representation in $\gdnf$
(Disjunctive Normal Form), $e=\bigvee _{i=1}^n\bigwedge _{j=1}^{n_i}s_{ij}$. We refer to the
set $\{S_1,\ldots ,S_n\}$, with $S_i=\{s_{i1},\ldots ,s_{in_i}\}$, as the set representation of $e$.
\end{definition}

In the following, we denote
by $K$ a set of $k$ directions $d_1,\ldots ,d_k$; by $N_P$ a set of primitive concepts;
by $x$ a $p$-ary spatial RA;
by $N_{cF}$ a finite set of concrete features referring to objects
in $\Delta _{\textsl{D}_x}$;
by $\alphabet$ the alphabet $2^{N_P}\times\Theta (N_{cF},\Delta
_{\textsl{D}_x})$,
$\Theta (N_{cF},\Delta _{\textsl{D}_x})$ being the set of total functions
$\theta :N_{cF}\rightarrow\Delta _{\textsl{D}_x}$, associating with each concrete feature $g$ a
concrete value $\theta (g)$ from the spatial concrete domain  $\Delta
_{\textsl{D}_x}$;
by $\lits (N_P)$ the set of literals derived from $N_P$ (viewed as a set of atomic
propositions):
$\lits (N_P)=N_P\cup\{\neg A:A\in N_P\}$;
by $c(2^{\lits (N_P)})$ the set of subsets of $\lits (N_P)$ which do
not contain a primitive concept and its negation: $c(2^{\lits
  (N_P)})=\{S\subset\lits (N_P):(\forall A\in N_P)(\{A,\neg
A\}\not\subseteq S)\}$;
by $\consts (x,K,N_{cF})$ the set of constraints of the form
$P(u_1,\ldots ,u_p)$ with $P$ being
an $x$ relation, $u_1,\ldots ,u_p$ $\ksgchains$ (i.e., $u_i$, $i\in\{1,\ldots ,p\}$, is of the form $g$ or $d_{i_1}\ldots d_{i_n}g$,
$n\geq 1$ and $n$ finite, the $d_{i_j}$'s being
directions in $K$, and $g$ a concrete feature).

\begin{definition}[$k$-ary $\Sigma$-tree]\label{karymtree}
Let $\Sigma$ and $K=\{d_1,\ldots ,d_k\}$, $k\geq 1$, be two disjoint alphabets: $\Sigma$ is a labelling alphabet and $K$ an
alphabet of directions. A (full) $k$-ary tree is an infinite tree
whose nodes $\alpha\in K^*$ have exactly $k$ immediate successors each,
$\alpha d_1,\ldots ,\alpha d_k$. A $\Sigma$-tree is a tree whose nodes are
labelled with elements of $\Sigma$. A (full) $k$-ary $\Sigma$-tree is
a $k$-ary tree $t$ which is also a $\Sigma$-tree, which we consider as
a mapping $t:K^*\rightarrow\Sigma$ associating with each node
$\alpha\in K^*$ an element $t(\alpha )\in\Sigma$. The empty word, $\epsilon$, denotes
the root of $t$. Given a node $\alpha\in K^*$ and a direction $d\in K$,
the concatenation of $\alpha$ and $d$, $\alpha d$, denotes the $d$-successor of
$\alpha$. The level $|\alpha |$ of a node $\alpha$ is the length of $\alpha$ as a word. We can
thus think of the edges of $t$ as being labelled with directions from $K$, and of the nodes of
$t$ as being labelled with letters from $\Sigma$. A partial $k$-ary
$\Sigma$-tree (over the set $K$ of directions) is a $\Sigma$-tree with
the property that a node may not have a $d$-successor for each
direction $d$; in other terms, a partial $k$-ary $\Sigma$-tree is a
$\Sigma$-tree which is a prefix-closed\footnote{$t$ is prefix-closed
  if, for all nodes $\alpha$, if $t$ is defined for $\alpha$ then it
is defined for all nodes $\alpha '$ consisting of prefixes of
  $\alpha$.} partial function $t:K^*\rightarrow\Sigma$.
\end{definition}

\begin{definition}[B\"uchi alternating automaton]\label{buechialtaut}
Let $k\geq 1$ be an integer and $K=\{d_1,\ldots ,d_k\}$ a set of directions.
An alternating automaton on $k$-ary $\alphabet$-trees is a tuple
$\textsl{A}=(\textsl{L}(\lits (N_P)\cup\consts (x,K,N_{cF})\cup K\times Q),
           \alphabet ,$ $\delta ,q_0,\textsl{F})$,
where
$Q$ is a finite set of states;
$\alphabet$ is the input alphabet (labelling the nodes of the input trees);
$\delta :Q\rightarrow \textsl{L}(\lits (N_P)\cup\consts (x,K,N_{cF})\cup K\times Q)$ is the
transition function;
$q_0\in Q$ is the initial state; and
$\textsl{F}$ is the set of accepting states.
\end{definition}
Let $\textsl{A}$ be an alternating automaton on $k$-ary
$\alphabet$-trees, as defined in Definition \ref{buechialtaut}, and $t$ a $k$-ary
$\alphabet$-tree. Given two alphabets $\Sigma _1$ and $\Sigma _2$, we
denote by $\Sigma _1\Sigma _2$ the concatenation of $\Sigma _1$ and
$\Sigma _2$, consisting of all words $ab$, with $a\in\Sigma _1$
and $b\in\Sigma _2$. In a run $r(\textsl{A},t)$ of $\textsl{A}$ on $t$ (see below),
which can be seen as an unfolding of a branch of the computation tree
$T(\textsl{A},t)$ of $\textsl{A}$ on $t$, as defined in
\cite{MullerS87a,MullerSS92a,MullerS95a}, the nodes of level $n$ will represent one possibility
for choices of $\textsl{A}$ up to level $n$ in $t$.
For each $n\geq 0$, we define the set of
$n$-histories to be the set
$H_n=\{q_0\}(KQ)^n$ of all $2n+1$-length words
consisting of $q_0$ as the first letter, followed by a $2n$-length
word $d_{i_1}q_{i_1}\ldots d_{i_n}q_{i_n}$, with $d_{i_j}\in K$ and $q_{i_j}\in Q$, for all
$j=1\ldots n$. If $h\in H_n$ and
$g\in KQ$ then $hg$, the
concatenation of $h$ and $g$, belongs to $H_{n+1}$. More generally, if $h\in H_n$ and
$e\in \textsl{L}(KQ)$, the
concatenation $he$ of $h$ and $e$ will denote the element of $\textsl{L}(H_{n+1})$ obtained by
prefixing $h$ to each generator in
$KQ$ which occurs in $e$.
Additionally, given an $n$-history $h=q_0d_{i_1}q_{i_1}\ldots d_{i_n}q_{i_n}$, with $n\geq 0$, we denote:
by $\last (h)$ the initial state $q_0$ if $h$ consists of the
    $0$-history $q_0$ ($n=0$),
and the state $q_{i_n}$ if $n\geq 1$;
by $\kproj (h)$ (the $K$-projection of $h$) the empty word
    $\epsilon$ if $n=0$, and the $n$-length word $d_{i_1}\ldots d_{i_n}$ otherwise; and
by $\qproj (h)$ (the $Q$-projection of $h$) the state $q_0$ if 
    $n=0$, and the $n+1$-length 
    word $q_0q_{i_1}\ldots q_{i_n}\in Q^{n+1}$ otherwise.
The union of all $H_n$, with $n$ finite, will be referred to as the
set of finite histories of $\textsl{A}$, and denoted by $\hf$. We denote by $\alphabett$ the alphabet
$2^{\hf}\times c(2^{\lits (N_P)})\times 2^{\consts (x,K,N_{cF})}$,
by $\alphabettt$ the alphabet $2^Q\times c(2^{\lits (N_P)})\times
2^{\consts (x,K,N_{cF})}$, and, in general, by $\alphabtt$ the
alphabet $S\times c(2^{\lits (N_P)})\times
2^{\consts (x,K,N_{cF})}$.

A run of the alternating automaton $\textsl{A}$ on $t$ is now introduced.
\begin{definition}[Run]\label{definitionrun}
Let $\textsl{A}$ be an alternating automaton on $k$-ary
$\alphabet$-trees, as defined in Definition \ref{buechialtaut}, and $t$ a $k$-ary
$\alphabet$-tree. A run, $r(\textsl{A},t)$, of $\textsl{A}$ on $t$ is a
partial $k$-ary $\alphabett$-tree defined
inductively as follows. For all directions $d\in K$, and for all nodes $u\in K^*$ of
$r(\textsl{A},t)$, $u$ has at most one outgoing edge labelled with $d$, 
and leading to the $d$-successor $ud$ of $u$. The label
$(Y_{\epsilon},L_{\epsilon},X_{\epsilon})$ of the root belongs to $2^{H_0}\times c(2^{\lits (N_P)})\times 2^{\consts (x,K,N_{cF})}$
---in other words, $Y_{\epsilon}=\{q_0\}$. If $u$ is a
node of $r(\textsl{A},t)$ of level $n\geq 0$, with label
$(Y_u,L_u,X_u)$, then calculate
$e=\bigwedge _{h\in Y_u}\distribute (h,\delta (\last (h)))$,
where $\distribute$ is a function associating with each pair $(h_1,e_1)$
of $\hf\times \textsl{L}(\lits (N_P)\cup\consts (x,K,N_{cF})\cup
K\times Q)$ an element of $\textsl{L}(\lits (N_P)\cup\consts
(x,K,N_{cF})\cup\hf)$ defined inductively in the following way:\\
$
\distribute (h_1,e_1)=
  \left\{
                   \begin{array}{l}
                         e_1
                               \mbox{ if }e_1\in \lits (N_P)\cup\consts (x,K,N_{cF}),  \\
                         h_1dq
                               \mbox{ if }e_1=(d,q)\mbox{, with
                                 }(d,q)\in K\times Q,  \\
                         \distribute (h_1,e_2)\vee\distribute (h_1,e_3)
                               \mbox{ if }e_1=e_2\vee e_3,  \\
                         \distribute (h_1,e_2)\wedge\distribute (h_1,e_3)
                               \mbox{ if }e_1=e_2\wedge e_3  \\
                   \end{array}
  \right.
$\\
Write $e$ in $\dnf$ as
$e=\bigvee _{i=1}^r(L_i\wedge X_i\wedge Y_i)$, where the $L_i$'s are
conjunctions of literals from $\lits (N_P)$, the $X_i$'s are
conjunctions of constraints from $\consts
(x,K,N_{cF})$, and the $Y_i$'s are
conjunctions of $n+1$-histories.
Then there exists $i=1\ldots r$ such that
\begin{enumerate}
  \item $L_u=\{\ell\in\lits (N_P):\ell\mbox{ occurs in }L_i\}$;
  \item $X_u=\{x\in\consts (x,K,N_{cF}):x\mbox{ occurs in }X_i\}$;
  \item for all $d\in K$, such that the set
$Y=\{hdq\in H_{n+1}: (h\in H_n)\mbox{ and }(q\in Q)\mbox{ and } (hdq\mbox{ occurs in }Y_i)\}$
is nonempty, and only for those $d$, $u$ has a $d$-successor, $ud$, whose label
$(Y_{ud},X_{ud},L_{ud})$ is such that
$Y_{ud}=Y$; and
  \item the label $t(u)=(\textsl{P}_u,\theta _u)\in 2^{N_P}\times\Theta (N_{cF},\Delta
_{\textsl{D}_x})$ of the node
$u$ of the input tree $t$ verifies the following, where, given a
        node $v$ in $t$, the notation $\theta _v$ consists of the function $\theta _v:N_{cF}\rightarrow\Delta _{\textsl{D}_x}$ which is the second
        argument of $t(v)$:
    \begin{enumerate}
      \item[$\bullet$] for all $A\in N_P$: if $A\in L_u$ then $A\in \textsl{P}_u$; and
        if $\neg A\in L_u$ then $A\notin \textsl{P}_u$ (the elements $A$ of
        $N_P$ such that, neither $A$ nor $\neg A$ occur in $L_u$, may or may not
        occur in $\textsl{P}_u$);
      \item[$\bullet$] for all $P(d_{1_1}\ldots d_{1_{n_1}}g_1,\ldots ,d_{p_1}\ldots d_{p_{n_p}}g_p)$ appearing in $X_u$,\\
        $P(\theta _{ud_{1_1}\ldots d_{1_{n_1}}}(g_1),\ldots ,\theta _{ud_{p_1}\ldots d_{p_{n_p}}}(g_p))$ holds. In other words, the values of 
        the concrete features $g_i$, $i\in\{1,\ldots ,p\}$, at the $d_{i_1}\ldots
        d_{i_{n_i}}$-successors of $u$ in $t$ are
        related by the $x$ relation $P$.
    \end{enumerate}
\end{enumerate}
A partial $k$-ary $\alphabett$-tree $\sigma$ is a run of $\textsl{A}$ if there exists a $k$-ary
$\alphabet$-tree $t$ such that $\sigma$ is a run of $\textsl{A}$ on $t$.
\end{definition}
\begin{definition}[CSP of a run]\label{cspofarun}
Let $\textsl{A}$ be an alternating automaton on $k$-ary
$\alphabet$-trees, as defined in Definition \ref{buechialtaut}, and $\sigma$ a run of
$\textsl{A}$:\\
(1) for all nodes $v$ of $\sigma$, of label
$\sigma (v)=(Y_v,L_v,X_v)\in 2^{\hf}\times c(2^{\lits (N_P)})\times
2^{\consts (x,K,N_{cF})}$, the argument $X_v$ gives rise to the CSP of $\sigma$ at $v$,
    $\csp _v(\sigma )$, whose set of variables, $V_v(\sigma )$, and set of constraints,
    $C_v(\sigma )$, are defined as follows:
(a) Initially, $V_v(\sigma )=\emptyset$ and $C_v(\sigma
      )=\emptyset$;
(b) for all $\ksgchains$ $d_{i_1}\ldots d_{i_n}g$ appearing in
      $X_v$, create, and add to $V_v(\sigma )$, a variable $\langle vd_{i_1}\ldots d_{i_n},g\rangle$;
(c) for all $P(d_{1_1}\ldots d_{1_{n_1}}g_1,\ldots ,d_{p_1}\ldots d_{p_{n_p}}g_p)$ in $X_v$, add the constraint\\
      $P(\langle vd_{1_1}\ldots d_{1_{n_1}},g_1\rangle,\ldots ,\langle vd_{p_1}\ldots d_{p_{n_p}},g_p\rangle)$ to $C_v(\sigma )$;\\
(2) the CSP of $\sigma$, $\csp (\sigma )$, is the CSP whose set of variables,
    $\textsl{V}(\sigma )$, and set of constraints, $\textsl{C}(\sigma )$, are defined
    as $\textsl{V}(\sigma )=\displaystyle\bigcup _{v\mbox{ node of }\sigma}V_v(\sigma )$ and
    $\textsl{C}(\sigma )=\displaystyle\bigcup _{v\mbox{ node of }\sigma}C_v(\sigma )$.
\end{definition}

An $n$-branch of a run $\sigma =r(\textsl{A},t)$ is a path of
length (number of edges) $n$ beginning at the root of
$\sigma$. A branch is an infinite path. If $u$ is the terminal node of an
$n$-branch $\beta$, then the argument $Y_u$ of the label $(Y_u,L_u,X_u)$ of
$u$ is a set of $n$-histories. Following \cite{MullerSS92a}, we say that
each $n$-history in $Y_u$ lies along $\beta$. An $n$-history $h$ lies
along $\sigma$ if there exists an $n$-branch $\beta$ of $\sigma$ such
that $h$ lies along $\beta$.
An (infinite) history is a sequence
$h=q_0d_{i_1}q_{i_1}\ldots d_{i_n}q_{i_n}\ldots\in\{q_0\}(KQ)^\omega$.
Given such a history,
$h=q_0d_{i_1}q_{i_1}\ldots d_{i_n}q_{i_n}\ldots\in\{q_0\}(KQ)^\omega$:
$h$ lies along a branch $\beta$ if, for every $n\geq 1$, the prefix of $h$
    consisting of the $n$-history
    $q_0d_{i_1}q_{i_1}\ldots d_{i_n}q_{i_n}$ lies along the
    $n$-branch $\beta _n$ consisting of the first $n$ edges of
    $\beta$;
$h$ lies along $\sigma$ if there exists a branch $\beta$ of
    $\sigma$ such that $h$ lies along $\beta$;
$\qproj (h)$ (the $Q$-projection of $h$) is the infinite word
    $q_0q_{i_1}\ldots q_{i_n}\ldots \in Q^\omega$ such that, for all $n\geq 1$, the
    $n+1$-length prefix $q_0q_{i_1}\ldots q_{i_n}$ is the $Q$-projection of
    $h_n$, the $n$-history which is the $2n+1$-prefix of $h$; and
we denote by $\infinity (h)$ the set of
states appearing infinitely often in $\qproj (h)$
The acceptance condition is now defined as follows. A history $h$ is accepting if
$\infinity (h)\cap \textsl{F}\not =\emptyset$. A branch
$\beta$ of $r(\textsl{A},t)$ is accepting if every history lying along
$\beta$ is accepting.

The condition for a run $\sigma$ to be accepting splits into two subconditions. The
first subcondition is the standard one, and is related to (the
histories lying along) the
branches of $\sigma$, all of which should be accepting. The second subcondition
is new: the CSP of $\sigma$, $\csp (\sigma )$,
should be consistent. $\textsl{A}$ accepts a $k$-ary $\alphabet$-tree $t$ if there exists an accepting
run of $\textsl{A}$ on $t$. The language $\textsl{L}(\textsl{A})$ accepted by $\textsl{A}$ is
the set of all $k$-ary $\alphabet$-trees accepted by $\textsl{A}$.

Informally, a run $\sigma =r(\textsl{A},t)$ is uniform if, for all $n\geq 0$, any two $n$-histories $h_1$ and $h_2$
 lying along $\sigma$, verifying $\kproj (h_1)=\kproj (h_2)$ ($n$-histories of a same node of $t$), and
suffixed (i.e., terminated) by the
same state, make the same transition. To define it formally, we
suppose that the transition function is such that $\delta (q)$, for all states $q$, is given as a
disjunction of conjunctions, in disjunctive normal form.

\begin{definition}[Uniform run]
Let $\textsl{A}$ be an alternating automaton on $k$-ary $\alphabet$-trees, as defined in
Definition \ref{buechialtaut}, and $\sigma$ a run of $\textsl{A}$. $\sigma$ is said to
be a uniform run $\iff$ it satisfies the following.
For all nodes $u$ of level $n\geq 0$, for all states $q$ in
$Q$, there exists a disjunct from $\delta (q)$, which we refer to as
$\delta (q,\sigma ,u)$, such that the following holds. Let
$(Y_u,L_u,X_u)$ be the label of $u$. Calculate
$e=\bigwedge _{h\in Y_u}\distribute (h,\delta (\last (h),\sigma ,u))$,
where $\distribute$ is defined as in Definition \ref{definitionrun}.
Write $e$ as
$e=L\wedge X\wedge Y$, where $L$ is a
conjunction of literals from $\lits (N_P)$, $X$ is a
conjunction of constraints from $\consts
(x,K,N_{cF})$, and $Y$ is a
conjunction of $n+1$-histories.
Then
%
$L_u=\{\ell\in\lits (N_P):\ell\mbox{ occurs in }L\}$;
%
$X_u=\{x\in\consts (x,K,N_{cF}):x\mbox{ occurs in }X\}$;
%
for all $d\in K$, such that the set
$Z=\{hdq\in H_{n+1}: (h\in H_n)\mbox{ and }(q\in Q)\mbox{ and }(hdq\mbox{ occurs in }Y)\}$
is nonempty, and only for those $d$, $u$ has a $d$-successor, $ud$, whose label
$(Y_{ud},X_{ud},L_{ud})$ is such that
$Y_{ud}=Z$.
\end{definition}

We show that the adding of a spatial concrete domain to alternating automata does not compromise the
uniformisation theorem for alternating automata
\cite{MullerSS92a,MullerS95a}. We then make use of the result to show that such an automaton can be simulated with a standard B\"uchi
nondeterministic automaton.
%
%
\begin{theorem}\label{urunthm}
Let $\textsl{A}$ be a B\"uchi alternating automaton on $k$-ary
$\alphabet$-trees, and $t$ a $k$-ary
$\alphabet$-tree. If $t$ is accepted by $\textsl{A}$ then there exists an accepting uniform run of $\textsl{A}$ on $t$.
\end{theorem}
%
%
{\bf Proof:} See additional material (separate file KR\_2018\_Supplement\_236). \cqfd
\section{B\"uchi automata on $k$-ary $\alphabet$-trees: simulating an alternating with a usual nondeterministic}
As a consequence of the uniformisation theorem (see its proof), a B\"uchi alternating automation on $k$-ary $\alphabet$-trees can be simulated with a standard
B\"uchi nondeterministic automaton on $k$-ary $\alphabet$-trees. Further background is needed before providing formally the simulating automaton.
\begin{definition}[distinguished levels of a branch]\label{distinguishedlevel}
Let $\textsl{A}$ be a B\"uchi alternating automaton, $\sigma$ a run of $\textsl{A}$, $\beta$ a branch of $\sigma$, and $\ell$ a positive integer.
$\ell$ is a distinguished level of $\beta$ $\iff$ there exists a sequence $n_0,\ldots ,n_k$ of positive integers verifying the following:
\begin{enumerate}
  \item $n_0<\ldots <n_k$
  \item $n_0$ is the smallest level such that each history lying along $\beta$ meets, from level $0$ to level $n_0$, at least once a state from $F$
  \item for all $i\geq 0$ such that $i<k$, $n_{i+1}$ is the smallest level such that:
    \begin{enumerate}
      \item $n_{i+1}\geq n_i+1$; and
      \item each history lying along $\beta$ meets, from level $n_i+1$ to level $n_{i+1}$, at least once a state from $F$
    \end{enumerate}
  \item $n_k=\ell$
\end{enumerate}
We use the notation $n^{\sigma ,\beta}_k$ to refer to such a distinguished level $\ell$: $\ell$ is the $k^{th}$ distinguished level of the branch $\beta$ of the run $\sigma$.
\end{definition}

Definition \ref{characterisingfunction} below makes use
of the integers (booleans) 0 and 1 to define a function referred to as the characterising function of a uniform run. The latter is then used by the characterising lemma,
Lemma \ref{characterisinglemma}, to characterise, and single out, the different distinguished levels of a branch of a uniform run.
The idea has been used in \cite{Isli93a,Isli96a} for the simulation of a B\"uch alternating automaton on infinite words with a B\"uchi nondeterministic automaton. We
adapt it to the simulation of a B\"uchi alternating automaton on $k$-ary $\alphabet$-trees.
\begin{definition}[characterising function]\label{characterisingfunction}
Let $\textsl{A}$ be a B\"uchi alternating automaton, and $\sigma$ a uniform run of $\textsl{A}$. The characterising function of $\sigma$, $u_{\sigma}$, is defined on the set of
$n$-histories lying along $\sigma$ as follows:
\begin{enumerate}
  \item
$
u_{\sigma}(q_0)=
  \left\{
                   \begin{array}{ll}
                         (q_0,1)
                               &\mbox{ if }q_0\in F,  \\
                         (q_0,0)
                               &\mbox{ otherwise}  \\
                   \end{array}
  \right.
$
  \item If $u_{\sigma}$ is known for all $n$-histories lying along $\sigma$, then it is defined as follows for an $n+1$-history $h=h'dq$:\\
	\underline{Case 1:} for all $n$-histories $h''$ lying along $\sigma$ and verifying $\kproj (h'')=\kproj (h')$, we have $u_{\sigma}(h'')\in Q\times\{1\}$:
          \begin{enumerate}
            \item[]
$
u_{\sigma}(h)=
  \left\{
                   \begin{array}{ll}
                         (q,1)
                               &\mbox{ if }q\in F,  \\
                         (q,0)
                               &\mbox{ otherwise}  \\
                   \end{array}
  \right.
$
          \end{enumerate}
	\underline{Case 2:} there are $n$-histories $h''$ lying along $\sigma$ and verifying $\kproj (h'')=\kproj (h')$, such that $u_{\sigma}(h'')\in Q\times\{0\}$:
          \begin{enumerate}
            \item[]
$
u_{\sigma}(h)=
  \left\{
                   \begin{array}{l}
                         (q,1)
                               \mbox{ if $q\in F$,  or all $n+1$-histories $h''dq$ }\\
                               \mbox{$\;\;\;\;\;\;$ verifying $\kproj (h')=\kproj (h'')$ are }\\
                               \mbox{$\;\;\;\;\;\;$ such that $u_{\sigma} (h'')\in Q\times\{1\}$,}\\
                         (q,0)
                               \mbox{ otherwise}  \\
                   \end{array}
  \right.
$
          \end{enumerate}
Given a node $v$ of $\sigma$ of level $n$, we denote by $E_{\sigma}(v)$ the set $\{u_{\sigma}(h):\mbox{ $h$ $n$-history lying along $\sigma$ verifying $\kproj (h)=v$}\}$.
\end{enumerate}
\end{definition}

\begin{lemma}[characterising lemma]\label{characterisinglemma}
Let $\textsl{A}$ be a B\"uchi alternating automaton, $\sigma$ a uniform run of $\textsl{A}$, $\beta$ a branch of $\sigma$, and $\ell$ a positive integer.
$\ell$ is a distinguished level of $\beta$ $\iff$ the node $v$ of $\beta$ of level $\ell$ verifies $E_{\sigma}(v)\subseteq Q\times\{1\}$
\end{lemma}

\begin{lemma}\label{lemmatwo}
Let $\textsl{A}$ be a B\"uchi alternating automaton, $\sigma$ a run of $\textsl{A}$, $\beta$ a branch of $\sigma$.
$\beta$ is accepting $\iff$ the number of its distinguished levels is infinite.
\end{lemma}

\begin{lemma}\label{lemmathree}
Let $\textsl{A}$ be a B\"uchi alternating automaton, $\sigma$ a uniform run of $\textsl{A}$, $\beta$ a branch of $\sigma$.
$\beta$ is accepting $\iff$ there exists a subset $Q_1$ of $Q\times\{1\}$ such that $\beta$ contains infinitely many nodes $v$ verifying
$E_{\sigma}(v)=Q_1$.
\end{lemma}

A B\"uchi nondeterministic automaton on $k$-ary $\alphabet$-trees can
be thought of as a special case of a B\"uchi alternating automaton: as 
one that sends, at each node of a run, in every direction, exactly one copy. In other
words, as a B\"uchi alternating automaton with the property that,
there is one and only one history lying along any branch of any run of the
automaton.
\begin{definition}[B\"uchi nondeterministic automaton]\label{buechinondetaut}
A B\"uchi nondeterministic automaton on $k$-ary $\alphabet$-trees is a tuple
$\textsl{B}=$ $(Q,K,\lits (N_P),\consts (x,K,N_{cF}),
           \alphabet ,\delta ,q_0,q_{\#},\textsl{F})$,
where
$Q$ is a finite set of states;
$K=\{d_1,\ldots ,d_k\}$ ($k\geq 1$) is a set of directions;
$\lits (N_P)$, $\consts (x,K,N_{cF})$ and
           $\alphabet$
are as in Definition \ref{buechialtaut};
$q_0\in Q$ is the initial state;
$\textsl{F}\subseteq Q$ defines the acceptance condition;
$q_{\#}\in F$ is an accept-all state, accepting all $k$-ary $\alphabet$-trees; and
$\delta :Q\rightarrow \textsl{P}(2^{\lits (N_P)}\times 2^{\consts
  (x,K,N_{cF})}\times Q^k)$ is the
transition function verifynig $\delta (q_{\#})=(\{\},\{\},(q_{\#},\cdots ,q_{\#}))$.
\end{definition}
\begin{definition}[Run of a B\"uchi nondeterministic automaton]
Let $\textsl{B}$ be a B\"uchi nondeterministic automaton on $k$-ary
$\alphabet$-trees, as defined in Definition \ref{buechinondetaut}, and $t$ a $k$-ary
$\alphabet$-tree. A run, $r (\textsl{B},t)$, of $\textsl{B}$ on $t$ is a
(full) $k$-ary $\alphabt$-tree\footnote{$\alphabt =Q\times c(2^{\lits (N_P)})\times
2^{\consts (x,K,N_{cF})}$.} defined
inductively as follows. For all directions $d\in K$, and for all nodes $u\in K^*$ of
$r(\textsl{B},t)$, $u$ has exactly one outgoing edge labelled with $d$, 
and leading to the $d$-successor $ud$ of $u$. The label
$(Y_{\epsilon},L_{\epsilon},X_{\epsilon})$ of the root belongs to $\{q_0\}\times c(2^{\lits (N_P)})\times 2^{\consts (x,K,N_{cF})}$
---in other words, $Y_{\epsilon}=q_0$. If $u$ is a
node of $r(\textsl{B},t)$ of level $n\geq 0$, with label
$(Y_u,L_u,X_u)$, then let
$e=\delta (Y_u)\subseteq 2^{\lits (N_P)}\times 2^{\consts
  (x,K,N_{cF})}\times Q^k$.
Then there exists $(L,X,(q_{i_1},\ldots ,q_{i_k}))\in\delta (Y_u)$ such that
$L_u=L$;
$X_u=X$;
for all $j=1\ldots k$, $u$ has a $d_j$-successor, $ud_j$, whose label
$(Y_{ud_j},X_{ud_j},L_{ud_j})$ is such that
$Y_{ud_j}=q_{i_j}$; and
the label $t(u)=(\textsl{P}_u,\theta _u)\in 2^{N_P}\times\Theta (N_{cF},\Delta
_{\textsl{D}_x})$ of the node
$u$ of the input tree $t$ verifies the following, where, given a
        node $v$ in $t$, the notation $\theta _v$ consists of the function $\theta _v:N_{cF}\rightarrow\Delta _{\textsl{D}_x}$ which is the second
        argument of $t(v)$:
    \begin{enumerate}
      \item[$\bullet$] for all $A\in N_P$: if $A\in L_u$ then $A\in \textsl{P}_u$; and
        if $\neg A\in L_u$ then $A\notin \textsl{P}_u$ (the elements $A$ of
        $N_P$ such that, neither $A$ nor $\neg A$ occur in $L_u$, may or may not
        occur in $\textsl{P}_u$);
      \item[$\bullet$] for all $P(d_{1_1}\ldots d_{1_{n_1}}g_1,\ldots ,d_{p_1}\ldots d_{p_{n_p}}g_p)$ appearing in $X_u$,\\
        $P(\theta _{ud_{1_1}\ldots d_{1_{n_1}}}(g_1),\ldots ,\theta _{ud_{p_1}\ldots d_{p_{n_p}}}(g_p))$ holds. In other words, the values of 
        the concrete features $g_i$, $i\in\{1,\ldots ,p\}$, at the $d_{i_1}\ldots
        d_{i_{n_i}}$-successors of $u$ in $t$ are
        related by the $x$ relation $P$.
    \end{enumerate}
A (full) $k$-ary $\alphabt$-tree $\sigma$ is a run of $\textsl{B}$ if there exists a (full) $k$-ary
$\alphabet$-tree $t$ such that $\sigma$ is a run of $\textsl{B}$ on $t$.
\end{definition}
\begin{definition}[CSP of a run]\label{cspofarun}
Let $\textsl{B}$ be a B\"uchi nondeterministic automaton on $k$-ary
$\alphabet$-trees, as defined in Definition \ref{buechinondetaut}, $t$ a $k$-ary
$\alphabet$-tree and $\sigma$ a run of
$\textsl{B}$ on $t$:
\begin{enumerate}
  \item for all nodes $v$ of $\sigma$, of label
$\sigma (v)=(Y_v,L_v,X_v)\in Q\times c(2^{\lits (N_P)})\times
2^{\consts (x,K,N_{cF})}$, the argument $X_v$ gives rise to the CSP of $\sigma$ at $v$,
    $\csp _v(\sigma )$, whose set of variables, $V_v(\sigma )$, and set of constraints,
    $C_v(\sigma )$, are defined as follows:
  \begin{enumerate}
    \item Initially, $V_v(\sigma )=\emptyset$ and $C_v(\sigma
      )=\emptyset$
    \item for all $\ksgchains$ $d_{i_1}\ldots d_{i_n}g$ appearing in
      $X_v$, create, and add to $V_v(\sigma )$, a variable $\langle vd_{i_1}\ldots d_{i_n},g\rangle$
    \item for all $P(d_{1_1}\ldots d_{1_{n_1}}g_1,\ldots ,d_{p_1}\ldots d_{p_{n_p}}g_p)$ in $X_v$, add the constraint\\
      $P(\langle vd_{1_1}\ldots d_{1_{n_1}},g_1\rangle,\ldots ,\langle vd_{p_1}\ldots d_{p_{n_p}},g_p\rangle)$ to $C_v(\sigma )$
  \end{enumerate}
  \item the CSP of $\sigma$, $\csp (\sigma )$, is the CSP whose set of variables,
    $\textsl{V}(\sigma )$, and set of constraints, $\textsl{C}(\sigma )$, are defined
    as $\textsl{V}(\sigma )=\displaystyle\bigcup _{v\mbox{ node of }\sigma}V_v(\sigma )$ and
    $\textsl{C}(\sigma )=\displaystyle\bigcup _{v\mbox{ node of }\sigma}C_v(\sigma )$.
\end{enumerate}
\end{definition}
An $n$-branch and a branch of a run of a B\"uchi nondeterministic
automaton are defined as in the alternating case. Given an $n$-branch
$\beta$, one and only one $n$-history lies along $\beta$, which is
$h=q_0d_{i_1}q_{i_1}\ldots d_{i_n}q_{i_n}\in\{q_0\}(KQ)^n$,
such that:
the node $\kproj (h)=d_{i_1}\ldots d_{i_n}$ is the terminal node of the $n$-branch; and
the label $(Y_u,X_u,L_u)$ of the $j$-th node $u=d_{i_1}\ldots d_{i_j}$ of the
  $n$-branch, $j=1\ldots n$, is such that $Y_u=q_{i_j}$.
An (infinite) history $h=q_0d_{i_1}q_{i_1}\ldots d_{i_n}q_{i_n}\ldots\in\{q_0\}(KQ)^\omega$
lies along a branch $\beta$ if, for every $n\geq 1$, the prefix of $h$
consisting of the $n$-history $q_0d_{i_1}q_{i_1}\ldots d_{i_n}q_{i_n}$ lies along the
$n$-branch $\beta _n$ consisting of the first $n$ edges of $\beta$. A history $h$ is accepting
if $\infinity (h)\cap\textsl{F}\not =\emptyset$. A branch is accepting if the history lying along it is
    accepting.

As in the alternating case, the condition for a run $\sigma$ to be accepting splits into two subconditions. The
first subcondition is the standard one, and is related to (the
histories lying along) the
branches of $\sigma$, all of which should be accepting. The second subcondition
is that the CSP of $\sigma$, $\csp (\sigma )$,
should be consistent. A B\"uchi nondeterministic automaton $\textsl{B}$ accepts a $k$-ary $\alphabet$-tree $t$ if there exists an accepting
run of $\textsl{B}$ on $t$. The language $\textsl{L}(\textsl{B})$ accepted by $\textsl{B}$ is
the set of all $k$-ary $\alphabet$-trees accepted by $\textsl{B}$.

The following corollary is a direct consequence of Theorem \ref{urunthm}.
\begin{corollary}\label{utheorem}
Let $\textsl{A}$ be a B\"uchi alternating automaton on $k$-ary
$\alphabet$-trees, $Q$ its set of states, and $F\subseteq Q$ its set of accepting states. There exists a B\"uchi nondeterministic automaton
simulating $\textsl{A}$, with a number of states bounded by $(\frac{2}{3})^{|F|}3^{|Q|}+1$,
the notation $|X|$, for a set $X$, standing for the cardinality of $X$.
\end{corollary}
{\bf Proof:} See additional material (separate file KR\_2018\_Supplement\_236). \cqfd

%
%
\section{The emptiness problem of a B\"uchi nondeterministic automaton on $k$-ary $\alphabet$-trees}
In the case of a standard (without constraints) B\"uchi nondeterministic automation on infinite words, the emptiness problem reduces to the existence,
in the automaton seen as a directed graph, of a strongly connected component  (reachable from the initial state and) containing an accepting state. In the
more general standard case of a B\"uchi nondeterministic automaton on (full) $k$-ary $\Sigma$-trees, the emptiness problem reduces to the existence of a small
(finite) tree model.

The uniformisation theorem for alternating automata on $k$-ary $\alphabet$-trees, Theorem \ref{urunthm}, allows to restrict attention, safely, to uniform runs.
This, in turn, as we have shown (Corollary \ref{utheorem}), means that such an automaton can be simulated with a nondeterministic one of the same type, on $k$-ary
$\alphabet$-trees. We give in this section an effective procedure generalising the finite tree model property to B\"uchi nondeterministic automata on $k$-ary $\alphabet$-trees.

A crucial point for the generalising procedure is the handling of the CSP of a run, which is potentially infinite. For the
purpose, we need another kind of a run, a regular run, which is
based on a function $\backconstraints$ :
\begin{enumerate}
  \item PTP is the acronym for ''Previously Targetted Parameters"
  \item the $\geq$ symbol means that the targetted parameters are reached at the current node or are not reached
           yet; in other words, the length of the path from the current node to the targetted parameters, in terms of number
           of nodes, is greater than or equal to $0$
\end{enumerate}
\begin{definition}[\backconstraints]
Let $\textsl{B}$ be a B\"uchi nondeterministic automaton on $k$-ary $\alphabet$-trees, $p$ the arity of the spatial RA $x$,
$\sigma$ a run of $\textsl{B}$ on an input $k$-ary $\alphabet$-tree $t$ and $u$ a node of $\sigma$. $\backconstraints (\sigma ,u)$
is defined inductively as follows:
\begin{enumerate}
  \item $\backconstraints (\sigma ,r)=\emptyset$, $r$ being the root of $\sigma$
  \item if $\backconstraints$ is known for node $v$ labelled with $(Q_1,L_1,C_1)$, then for the immediate $d$-successor $v'=vd$
of $v$, $d\in\{d_1,\ldots ,d_k\}$, $\backconstraints (\sigma ,v')$ is defined as follows:
  \item initialise $\backconstraints (\sigma ,v')$ to the empty set: $\backconstraints (\sigma ,v')=\emptyset$
  \item for all $i=1$ to $p$
    \begin{enumerate}
      \item For all constraints $c$ of the form $P(u_1,\ldots ,u_{i-1},du_i,u_{i+1},\ldots ,u_p)$ appearing in $C_1$:
		$\backconstraints (\sigma ,v')=\backconstraints (\sigma ,v')\cup\{(c,i,u_i)\}$
      \item For all triples $(c,i,du)$ appearing in $\backconstraints (\sigma ,v)$:
		$\backconstraints (\sigma ,v')=\backconstraints (\sigma ,v')\cup\{(c,i,u)\}$
    \end{enumerate}
\end{enumerate}
\end{definition}
If $\backconstraints (\sigma ,v)$ contains the triple $(c,i,u)$, this means the following:
the constraint $c$ is of the form $P(u_1,\ldots ,u_i,\ldots ,u_p)$;
$u_i$ is of the form $d_{i_1}\ldots d_{i_m}u$;
$v$ is of the form $wd_{i_1}\ldots d_{i_m}$;
the label $(Q_1,L_1,C_1)$ of node $w$ verifies $c\in C_1$;
furthermore, $u$ is of either form $g$ or $d_{i_{m+1}}\ldots d_{i_{\ell}}g$:
    \begin{enumerate}
      \item if $u$ is of the form $g$ then the parameter targetted by the constraint $c$ of the $d_{i_1}\ldots d_{i_m}$-predecessor $w$
           of $v$, is the concrete feature $g$ at the current node $v$ of the input tree $t$
      \item if $u$ is of the form $d_{i_{m+1}}\ldots d_{i_{\ell}}g$ then the parameter targetted by the constraint $c$, is the concrete
               feature $g$ at the $d_{i_{m+1}}\ldots d_{i_{\ell}}$-successor of $v$ of the input tree $t$
    \end{enumerate}


\begin{definition}[prefix and lexicographic order]
Let $\Sigma =\{a_1,\ldots ,a_n\}$ be an ordered alphabet, with $a_1<a_2<\cdots <a_n$, and
$u,v\in\Sigma ^*$. The relations ``$u$ is prefix of $v$'', denoted by $\prefix (u,v)$,
and ``$u$ is lexicographically smaller than $v$'', denoted by $u\lleq v$, are defined in
the following obvious manner:
$\prefix (u,v)$ $\iff$ $v=uw$, for some $w\in\Sigma ^*$;
$u\lleq v$ $\iff$, either $\prefix (u,v)$, or $u=w_1aw_2$ and $v=w_1bw_3$, for
    some $w_1,w_2,w_3\in\Sigma ^*$ and $a,b\in\Sigma$, with $a<b$.
\end{definition}
We will also need the derived relations ``$u$ is a strict prefix of $v$'', ``$u$ is
lexicographically strictly smaller than $v$'', and ``$u$ and $v$ are incomparable'',
which we denote, respectively, by $\sprefix (u,v)$, $u\slleq v$ and
$\incomparable (u,v)$:
$\sprefix (u,v)$ $\iff$ $\prefix (u,v)$ and $u\not = v$;
$u\slleq v$ $\iff$ $u\lleq v$ and $u\not = v$;
$\incomparable (u,v)$ $\iff$ $\neg\prefix (u,v)$ and $\neg\prefix (v,u)$.
%
\begin{definition}[subtree]
Let $K=\{d_1,\ldots ,d_k\}$ be a set of $k$ directions, $t$ a
partial $k$-ary $\Sigma$-tree,
and $u\in K^*$ a node of $t$. The subtree of $t$ at $u$, denoted $t/u$, is the
partial $k$-ary $\Sigma$-tree $t'$, whose nodes are of the form $v$, so that $uv$ is a 
node of $t$, and, for all such nodes, $t'(v)=t(uv)$ ---i.e., the
label of $v$ in $t'$, is the same as the one of $uv$ in $t$.
\end{definition}
\begin{definition}[substitution]
Let $K=\{d_1,\ldots ,d_k\}$ be a set of $k$ directions, $t$ and $t'$ two
partial $k$-ary $\Sigma$-trees, and
$u\in K^*$ a node of $t$.
The substitution of $t'$ to the subtree of $t$ at $u$, or $u$-substitution of $t'$
in $t$, denoted $t(u\leftarrow t')$, is the partial $k$-ary $\Sigma$-tree $t''$ such
that, the nodes are of the form $v$, with $v$ node of $t$ of which $u$ 
is not a prefix, or of the form $uv$, with $v$ a node of $t'$. The
label $t''(v)$ of $v$ in $t''$ is defined as follows:
$
t''(v)=
  \left\{
                   \begin{array}{ll}
                         t'(w)
                               &\mbox{ if }v=uw\mbox{, for some node
                                 }w\mbox{ of }t',  \\
                         t(v)
                               &\mbox{ otherwise}  \\
                   \end{array}
  \right.
$
\end{definition}
\begin{definition}[cut]
Let $K=\{d_1,\ldots ,d_k\}$ be a set of $k$ directions, $t$ a
partial $k$-ary $\Sigma$-tree, and
$u\in K^*$ a node of $t$.
The cut in $t$ of the subtree at $u$, or $u$-cut
in $t$, denoted $c(u,t)$, is the partial $k$-ary $\Sigma$-tree $t'$
whose nodes are those nodes $v$ of $t$ of which $u$  
is not a strict prefix ---i.e., such that $\neg\sprefix (u,v)$. The
label $t'(v)$ of any node $v$ in $t'$ is the same as $t(v)$, the label of the same node in $t$.
\end{definition}
%
%
\begin{figure}[t]
\begin{scriptsize}
\begin{enumerate}
  \item {\bf Input:} an accepting run $\sigma$ of a B\"uchi nondeterministic automaton $\textsl{B}$.
  \item {\bf Output:} a finite tree $s_{\sigma}$ obtained from $\sigma$, from which a regular
    run of $\textsl{B}$ can be generated.
  \item\label{lone} Initialise $s_{\sigma}$ to $\sigma$: $s_{\sigma}\leftarrow\sigma$;
  \item\label{ltwo} Initially, no node of $s_{\sigma}$ is marked;
  \item\label{lthree} {\bf repeat} while possible\{
  \item\label{lfour} \hskip 0.2cm Let $v$ be the smallest non marked node of $s_{\sigma}$ such that there exists a 
    non marked node $u$ of $s_{\sigma}$, so that $\slleq (u,v)$
    \underline{and} $Y_u=Y_v$ \underline{and} $\backconstraints (\sigma 
    ,u)=\backconstraints (\sigma ,v)$;
  \item\label{lsix} \hskip 0.2cm if $\neg\prefix (u,v)$\{
  \item\label{lseven} \hskip 0.6cm $s_{\sigma}\leftarrow c(v,s_{\sigma})$; $\backnode (v)\leftarrow u$; mark $v$;
  \item\label{lten} \hskip 0.6cm \}
  \item\label{leleven} \hskip 0.2cm else \underline{\% $\prefix (u,v)$ \%}
  \item\label{ltwelve} \hskip 0.6cm if there exists a node $w$ between $u$ and $v$ (i.e., so that $\lleq (u,w)\wedge\lleq (w,v)$) verifying
    $Y_w\in\textsl{F}$
    then\{
%
%
  \item\label{lthirteen} \hskip 1cm $s_{\sigma}\leftarrow c(v,s_{\sigma})$; $\backnode (v)\leftarrow u$; mark $v$;
  \item\label{lsixteen} \hskip 1cm \}
  \item\label{lseventeen} \hskip 0.6cm else\{
  \item\label{leighteen} \hskip 1cm $s'\leftarrow s_{\sigma}/v$; $s_{\sigma}\leftarrow s_{\sigma}(u\leftarrow s')$;
  \item\label{ltwenty} \hskip 1cm \}
  \item\label{ltwentyone} \hskip 0.2cm \} \underline{\% end {\bf repeat} \%}
\end{enumerate}
\caption{The order $d_1<\ldots <d_k$ is assumed on the directions in $K$.}\label{finitekarytree}
\end{scriptsize}
\end{figure}
\begin{definition}[regular run]
Let $\textsl{B}$ be a B\"uchi nondeterministic automaton on $k$-ary
$\alphabet$-trees, as defined in Definition \ref{buechialtaut}, and
$\sigma$ a run of $\textsl{B}$. $\sigma$ is regular if,
for all nodes $u$ and $v$ of $\sigma$ verifying $\backconstraints
(\sigma ,u)=\backconstraints (\sigma ,v)$, and whose labels $\sigma
(u)=(Y_u,L_u,X_u)$ and $\sigma
(v)=(Y_v,L_v,X_v)$ verify $Y_u=Y_v$, the following holds:
$L_u=L_v$;
$X_u=X_v$; and
for all $d\in K$, it is the case that $Y_{ud}=Y_{vd}$.
\end{definition}
\begin{theorem}\label{gafrunthm}
Let $\textsl{B}$ be a B\"uchi nondeterministic automaton on $k$-ary
$\alphabet$-trees. There exists an accepting run of $\textsl{B}$
$\iff$ there exists an accepting regular run of $\textsl{B}$.
\end{theorem}
{\bf Proof:} A regular run is a particular case of a run, which 
means that the existence of an accepting regular run
implies the existence of an accepting run. To show the other
direction of the equivalence, suppose the existence of an accepting
run, say $\sigma$. From $\sigma$, we first build a finite partial
$k$-ary $\alphabt$-tree, $s_{\sigma}$. We then show how to use $s_{\sigma}$ to get an
accepting regular run of $\textsl{B}$. The tree $s_{\sigma}$ is built by the
procedure of Figure \ref{finitekarytree}. There are three key points in the procedure :\\
%
{\bf First key point:} if $u$ is not prefix of $v$: given that
    $Y_u=Y_v$ and $\backconstraints (\sigma ,u)=\backconstraints (\sigma ,v)$, we
    can substitute the subtree of $s_{\sigma}$ at $u$ to the subtree of $s_{\sigma}$ at
    $v$, and get a run with all branches accepting, and with a global CSP
    consistent. The procedure, however, does not do the
    substitution. Instead, it cuts the subtree at $v$, marks $v$ and sets $u$
    as the back node of $v$, information which will be used in the
    building of the accepting regular run (line
    \ref{lseven}).\\
If $u$ is a (strict) prefix of $v$ then there are
    two possibilities:
    \begin{enumerate}
      \item[\#] {\bf Second key point:} if there exists a node $w$ between $u$ and $v$ so that
        $Y_w\in\textsl{F}$ (line
        \ref{ltwelve}) then cutting $s_{\sigma}$ at $v$, and then repeatedly pasting 
        the subtree at $u$ of the resulting tree, will lead to an
        accepting run, again thanks to $\backconstraints (\sigma
        ,u)=\backconstraints (\sigma ,v)$. What the procedure does in
        this case: it cuts the subtree at $v$,  marks $v$ and sets $u$
    as the back node of $v$, information which will be used in the
    building of the accepting regular run  (line
    \ref{lthirteen})
      \item[\#] {\bf Third key point:} the other possibility corresponds to the case when the 
        segment $[u,v]$ does not contain nodes $w$ verifying $Y_w\in\textsl{F}$. The procedure shortens the way to segments
        $[u,v]$ including nodes $w$ verifying $Y_w\in\textsl{F}$, by substituting the subtree at $v$ to the subtree at $u$ (line
    \ref{leighteen}). The repetition of this shortening will eventually lead at some point to a segment $[u,v]$
    with the requirement $\backconstraints (\sigma ,u)=\backconstraints (\sigma ,v)$ and including nodes $w$ verifying $Y_w\in\textsl{F}$, due to finiteness of the cross-product ...
    In particular, given that the input run is accepting, each of its branches is such that there exists a state $q\in\textsl{F}$ and
    two distinct nodes $s_1$ and $s_2$ of the branch, so that $Y_{s_1}=Y_{s_2}=q$ and 
    $\backconstraints (\sigma ,s_1)=\backconstraints (\sigma ,s_2)$.
    \end{enumerate}
The output tree $s_{\sigma}$ of the procedure of Figure \ref{finitekarytree} is clearly finite. To see it, suppose that it's not. This would mean
that $s_{\sigma}$ has an infinite branch, say $\beta$. Given that the run we started with is accepting, $\beta$ would repeat infinitely often an
element $q$ of $F$, and therefore would contain infinitely many nodes $u_i$, $i\geq 1$, such that $Y_{u_i}=q$, for all $i\geq 1$,
and $\backconstraints (\sigma ,u_i)=\backconstraints (\sigma ,u_j)$, for all $i,j$. A simple look at the three key points suffices to
see that this would lead to a contradiction.

Furthermore, the tree $s_{\sigma}$ verifies the following:
\begin{enumerate}
  \item the marked nodes of $s_{\sigma}$ are exactly its leaves;
  \item each leaf node $v$ of $s_{\sigma}$ is is such that, there is one and only one internal node $u$ of $s_{\sigma}$ verifying
$Y_u=Y_v$ and $\backconstraints (\sigma ,u)=\backconstraints (\sigma ,v)$. Furthermore, either $u$ is a (strict) prefix of $v$
and there exists a node $w$ between $u$ and $v$ verifying $Y_w\in F$; or, $\lleq (u,v)$ but $u$ is not a prefix of $v$. For
each such node $v$, we refer to the corresponding internal node $u$ as $i_v$, and to the subtree of $s_{\sigma}$ at $u$ as $s_{\sigma}/i_v$.
\end{enumerate}
%
%
From $s_{\sigma}$, we now build an accepting regular run $\sigma '$ by, intuitively, initialising $\sigma '$ to $s_{\sigma}$, and then
repeating the process of pasting at a leaf node $v_1$ of $\sigma '$ a subtree $t'$ of $s_{\sigma}$ whose root matches $v_1$:
\begin{scriptsize}
\begin{enumerate}
  \item[] {\bf Step $0$:}
  \item $\sigma _0\leftarrow s_{\sigma}$
  \item[] {\bf Step $1$:}
  \item  \hskip 0.3cm initialise $\sigma _1$ to $\sigma _0$: $\sigma
    _1\leftarrow\sigma _0$
  \item  \hskip 0.3cm repeat while possible\{
    \begin{enumerate}
      \item let $v_1$ be a leaf node of $\sigma _1$ of level $1$
      \item let $v_1'$ be the leaf node of $s_{\sigma}$ of which $v_1$ is a copy
      \item let $v_2$ be the unique internal node of $s_{\sigma}$ verifying $Y_{v_2}=Y_{v_1'}$ and 
              $\backconstraints (s_{\sigma},v_2)=\backconstraints (s_{\sigma},v_1')$
       \item $\sigma _1\leftarrow\sigma _1(v_1\leftarrow t/v_2)$
       \item[\}]
    \end{enumerate}
  \item[] {\bf Step $n$ $(n\geq 2)$:}
  \item initialise $\sigma _n$ to $\sigma _{n-1}$: $\sigma _n\leftarrow\sigma _{n-1}$
  \item repeat while possible\{
    \begin{enumerate}
      \item let $v_1$ be a leaf node of $\sigma _n$ of level $n$
      \item let $v_1'$ be the leaf node of $s_{\sigma}$ of which $v_1$ is a copy
      \item let $v_2$ be the unique internal node of $s_{\sigma}$ verifying $Y_{v_2}=Y_{v_1'}$ and 
              $\backconstraints (t,v_2)=\backconstraints (t,v_1')$
       \item $\sigma _n\leftarrow\sigma _n(v_1\leftarrow t/v_2)$
       \item[\}]
    \end{enumerate}
\end{enumerate}
\end{scriptsize}
%
%
Clearly, $\lim\limits_{
\begin{array}{l} 
n \to +\infty
\end{array}}\sigma _n$, the limit of $\sigma _n$ when $n$ tends to $+\infty$, is an accepting regular run. \cqfd

The first corollary gives a polynomial bound on the size of $s_{\sigma}$ in terms of number of nodes.

\begin{corollary}\label{xdlsatisfiability1}Let $\textsl{B}$ be a B\"uchi nondeterministic automaton on $k$-ary $\alphabet$-trees, $p$ the arity of the spatial RA $x$,
$\ell _{fc}$ and $n_c$, respectively, the length of the longest $\ksgchain$ and the number of
constraints from $\consts (x,K,N_{cF})$ appearing in the transition
function $\delta$ of $\textsl{B}$,
$\sigma$ a run of $\textsl{B}$ on an input $k$-ary $\alphabet$-tree $t$. The
number of internal nodes and the number of leaf nodes of $s_{\sigma}$ are bounded by $|Q|\times n_c\times\ell _{fc}\times p$ and $|Q|\times n_c\times\ell _{fc}\times p\times k$, respectively. \cqfd
\end{corollary}

The following corollary is a direct consequence of Theorem \ref{gafrunthm}.
\begin{corollary}\label{xdlsatisfiability2}
There exists a nondeterministic doubly depth-first polynomial space algorithm deciding
the emptiness problem of a B\"uchi nondeterministic automaton on $k$-ary $\alphabet$-trees.
\end{corollary}
{\bf Proof:} Let $\textsl{B}$ be a B\"uchi nondeterministic automaton on $k$-ary $\alphabet$-trees,
$\sigma$ a run of $\textsl{B}$ on an input $k$-ary $\alphabet$-tree $t$, and $p$,
$\ell _{fc}$ and $n_c$ as in Corollary \ref{xdlsatisfiability1}. The number of nodes of the output tree $s_{\sigma}$ of
the procedure of Figure \ref{finitekarytree}, is polynomially bounded by
$|Q|\times n_c\times\ell _{fc}\times p\times(k+1)$. We can thus
build such a tree, if it exists, or report its inexistence, otherwise, using a nondeterministic doubly depth-first
polynomial space algorithm :
\begin{enumerate}
  \item "or" depth-first: the classical notion of depth-first, which will govern here the choices offered by the transition function ("or" branching)
  \item "and" depth-first: the construction of the finite-tree representation of an accepting regular run we are looking for, if any, is done in a
depth-first manner, according to the order $d_1<\ldots <d_k$ on the set of directions ("and" branching).
\end{enumerate}

\begin{figure}
\begin{scriptsize}
main()\{
\begin{enumerate}
  \item global variables : $ftm\_root$, $gcsp\_checked$
  \item\label{main1} $ftm\_root\leftarrow create\_node(TREE)$; $ftm\_root.state\leftarrow q_0$;
  \item\label{main3} $ftm\_root.backnode\leftarrow -1$; $ftm\_root.PTPge\leftarrow\emptyset$;
  \item\label{main42} $ftm\_root.word\leftarrow ""$; $gcsp\_checked\leftarrow false$;
  \item\label{main5} {\bf if} ftm($ftm\_root$)\{print("not-empty"); print\_tree($ftm\_root$)\}
  \item\label{main6}  {\bf else} print("empty") \underline{\% endif \%}
  \item[\}] \underline{\% end main \%}
\end{enumerate}
function $globalcsp(s)$\{
\begin{enumerate}
          \item variables local to the function: $result$, $C\_set$, $C\_cst$
          \item $result\leftarrow\emptyset$;
          \item {\bf if} $s.1\neq nil$\{ \underline{\% $s$ internal node \%}
            \begin{enumerate}
              \item $C\_set\leftarrow s.constraints$;
              \item {\bf while} $C\_set\neq\emptyset$\{
                \begin{enumerate}
                  \item let $C\_cst$ be an element of $C\_set$; \underline{\% $C\_cst$ of the form $P(u_1,\ldots ,u_r)$ \%}
                  \item $result\leftarrow result\cup\{P(variable(s,u_1),\ldots ,variable(s,u_r))\}$
                  \item $C\_set\leftarrow C\_set\setminus\{C\_cst\}$\}
                \end{enumerate}
                  \item {\bf for} i=1 {\bf to} $k$\{$result\leftarrow result\cup globalcsp(s.i);$\}\}
            \end{enumerate}
          \item return $result$;\}
\end{enumerate}
function $variable(s,u)$\{
\begin{enumerate}
          \item\label{ftm12} {\bf if} $s.1= nil$\{ \underline{\% $s$ leaf node \%}
            \begin{enumerate} 
              \item\label{ftm12} return $variable(s.backnode,u)$\}
            \end{enumerate} 
          \item\label{ftm15} {\bf else}
            \begin{enumerate}
              \item\label{ftm17} {\bf if} ($u$ is the concrete feature $g$)\{return $<s,g>$;\}
              \item\label{ftm17} {\bf else}\{ \underline{\% $u$ of the form $du'$, with $d\in K$ \%}
                \begin{enumerate} 
                  \item\label{ftm12} return $variable(s.d,u')$\}\}
                \end{enumerate}
            \end{enumerate}
\end{enumerate}
\caption{a not-empty/empty answer to the emptiness problem of a B\"uchi nondeterministic automaton $\textsl{B}=(Q,K,\lits (N_P),\consts (x,K,N_{cF}),
           \alphabet ,\delta ,q_0,q_{\#},\textsl{F})$, with, in the case of a not-empty answer,
	   a finite tree representation of an accepting regular run of $\textsl{B}$.}\label{ftmmain}
\end{scriptsize}
\end{figure}

\begin{figure}
\begin{scriptsize}
procedure ftm(s) \underline{\% ftm for "finite tree model" \%}
\begin{enumerate}
  \item variables local to the procedure : $Delta$, $j$, $backtrack$, $s'$, $S$
  \item\label{ftm1} $\mbox{Delta}\leftarrow\delta (s.state)$; \hskip 0.5cm \% Delta of the from $\{S_1,\ldots ,S_r\}$ \%
  \item\label{ftm2} {\bf for} i=1 {\bf to} $r$\{ \underline{\% or branching \%}
    \begin{enumerate}
      \item\label{ftm3} $OrBr\leftarrow S_i$; \underline{\% OrBr of the from $[L_1,C_1,(q_1,\ldots ,q_k)]$ \%}
      \item\label{ftm4} $s.constraints\leftarrow C_1$;
      \item\label{ftm5} $j\leftarrow 1$; $\mbox{backtrack}\leftarrow false$;
      \item\label{ftm6} {\bf while} $j\leq k$ {\bf and} $\neg\mbox{backtrack}$\{ \underline{\% and branching \%}
        \begin{enumerate}
          \item\label{ftm7} $s'\leftarrow create\_node(TREE)$;
          \item\label{ftm8} $s.j\leftarrow s'$; $s'.state\leftarrow q_j$; $s'.word\leftarrow concatenate(s.word,"d_j")$;
          \item\label{ftm9} initialise $s'.backnode$ to -1: $s'.backnode\leftarrow -1$;
          \item\label{ftm10} $s'.PTPge\leftarrow\backconstraints (s')$;
          \item\label{ftm11} $S=\{s'':\mbox{ ($s''$ internal node) and }\slleq (s''.word,s'.word)\mbox{ and }s''.state=s'.state\mbox{ and }s''.PTPge=s'.PTPge\}$;
          \item\label{ftm12} {\bf if} $S=\emptyset$ {\bf then} $backtrack\leftarrow\neg ftm(s')$
          \item\label{ftm13} {\bf else}\{
            \begin{enumerate}
              \item\label{ftm14} let $s''$ be the unique element of $S$;
              \item\label{ftm15} {\bf if} $ThirdKeyPoint(s'',s')$ {\bf then} $backtrack\leftarrow true$
              \item[] {\bf else}\{
              \item\label{ftm17} \hskip 1cm $s'.backnode\leftarrow s''$
              \item\label{ftm18} \hskip 1cm {\bf for} i=1 {\bf to} $k$\{$s'.j\leftarrow nil;$\} \underline{\% mark $s'$ as a leaf node \%}
              \item[] \hskip 1cm \} \underline{\% end internal if \%}
              \item[\}] \underline{\% end external if \%}
            \end{enumerate}
          \item\label{ftm19} $j\leftarrow j+1$
          \item[\}] \underline{\% endwhile \%}  
    \end{enumerate}
      \item {\bf if} $backtrack$\{\underline{\% restore : cut the subtree at $s$ \%}
        \begin{enumerate}
          \item\label{ftm20} $ftm\_root\leftarrow c(ftm\_root,s)$;
          \item[\}]
        \end{enumerate}
      \item\label{ftm21} {\bf else}\{ \underline{\% a subtree at $s$ successfully built \%}
        \begin{enumerate}
          \item\label{ftm22} {\bf if}($s.word\in\{d_k\}^*$) \underline{\%  a whole tree successfully built \%}


          \item\label{ftm222} {\bf if}($gcsp\_checked$)\{return true;\} \underline{\%  consistency of the global CSP already checked \%}
          \item\label{ftm223} {\bf else}\{ \underline{\% check consistency of the global CSP \%}

            \begin{enumerate}
              \item\label{ftm23} $csp.constraints\leftarrow globalcsp(ftm\_root)$;
              \item\label{ftm232} $csp.variables\leftarrow\{<s,g>:\mbox{ }<s,g>\mbox{ occurs in }csp.constraints\}$;
              \item\label{ftm24} {\bf if} consistent(csp)\{$gcsp\_checked\leftarrow true$; return true;\}
              \item[] {\bf else}\{ \underline{\% restore: cut the subtree at $s$ \%}
              \item\label{ftm242} \hskip 0.5cm $ftm\_root\leftarrow c(ftm\_root,s)$;
              \item[] \hskip 0.5cm\} \underline{\% end most internal if-else \%}
              \item[\}] \underline{\% end second most internal if-else \%}
            \end{enumerate}


              \item {\bf else}\{\underline{\%  continue the attempt to build a whole tree \%}
                \begin{enumerate}
                  \item\label{ftm26} return true;
                  \item[\}] \underline{\% end second most external if-else \%}
                \end{enumerate}
              \item[\}]\underline{\% end most external if-else \%}
        \end{enumerate}
      \item[\}] \underline{\% endfor \%}
    \end{enumerate}
  \item\label{ftm27} {\bf return} false; \underline{\% failure to expand the tree beyond node $s$ \%}
  \item[\}] \underline{\% end ftm \%}
\end{enumerate}
\caption{The procedure $ftm$.}\label{ftm}
\end{scriptsize}
\end{figure}

Figure \ref{ftm} presents such an algorithm as a procedure $ftm$ (finite tree model). A main program (Figure \ref{ftmmain}) launches the construction of
the finite tree representation. A data type named $TREE$ is used, consisting of a record type with the following fields:
(1) the label of a node $s$ of a run $\sigma$ of $\textsl{B}$ belongs to $\alphabt =Q\times c(2^{\lits (N_P)})\times 2^{\consts (x,K,N_{cF})}$,
	and is of the form $(q_1,L_1,C_1)$: two fields of TREE, $state$ and $constraints$, are used to store, for each node $s$, the components $q_1$
	and $C_1$ of its label;
(2) a third field, $PTPge$, will record, for each node $s$, the set of "Previously Targetted Parameters" $\backconstraints (s)$;
(3) a fourth field, $backnode$, is used to store, for each leaf node $s$, the node $\backnode (s)$ (for internal nodes, the field $backnode$ is set to $-1$);
(4) a fifth field, $word$, gives, for each node $s$, its representation as a word of $K^*$;
(5) finally, $k$ other fields, named $1, 2,\ldots ,k$, are used to store, for each node $s$, pointers to the $k$ immediate successors of $s$, one per direction in $K$.
The initialisation of the finite tree construction, at the main program, creates the root of the tree as a record of type $TREE$, for which the fields $state$, $backnode$,
$PTPge$ and $word$ are set to $q_0$, $-1$, $\emptyset$ and $""$, respectively (the representation of the root as a word of $K^*$ is the empty word $""$).
The root is pointed at by the pointer variable $ftm\_root$. The program then calls the recursive
procedure $ftm$ to finish the construction, if (a finite tree representation of) an accepting regular run exists, or to report a yes answer to the emptiness problem of
$\textsl{B}$, otherwise.

The call $ftm(s)$ of the recursive procedure $ftm$ aims at expanding the tree beyond node $s$. The possible ways of expanding node $s$ are in the choices offered
by $\delta (s.state)=\{S_1,\ldots ,S_r\}$. The choice $S_i$, which is of the
form $[L_1,C_1,(q_1,\ldots ,q_k)]$, a triple of $2^{\lits (N_P)}\times 2^{\consts (x,K,N_{cF})}\times Q^k$, is used as follows in an attempt to expand the tree beyond
$s$. First, the field $constraints$ of $s$ is set to $C_1$. Then the expansion will succeed $\iff$ the and branching succeeds in each of the $k$ directions, $d_1,\ldots ,d_k$.
The conditions for success of the and branching in the $j^{th}$ direction, $j\in\{1,\ldots ,k\}$, are the following:
(1) a node $s'$ is created as the $j^{th}$ successor of $s$;
(2) the fields $state$, $backnode$, $PTPge$ and $word$ of $s'$ are set to $q_j$, $-1$, $\backconstraints (s')$ and $concatenate(s.word,"d_j")$, respectively;
(3) the three key points discussed earlier are then looked at to decide the following points:\\
(A) if there exists an internal node $s''$ such that $s''.state=s'.state$ and $s''.PTPge=s'.PTPge$ then:
        \begin{enumerate}
          \item the and branching fails in the $j^{th}$ direction if the third key point applies to nodes $s''$ and $s'$: this correponds to $\sprefix (s''.word,s'.word)$ and there is no node $s'''$ such that $\prefix (s'',s''')$ and $\prefix (s''',s')$ and $s'''.state\in F$: the current choice of $\delta (s.state)$ fails, the subtree at
                      $s$ is cut from the tree being constructed, and the procedure jumps to the next choice of $\delta (s.state)$, if any, or expresses its failure to expand the
                      tree beyond $s$ by returning $false$ (Figure \ref{ftm}: lines \ref{ftm15}, \ref{ftm20} and \ref{ftm27})
          \item it succeeds otherwise (i.e., if either of the other two key points applies to $s''$ and $s'$): this correponds to $\slleq (s''.word,s'.word)$ and $\neg\sprefix (s''.word,s'.word)$, or $\sprefix (s''.word,s'.word)$ and there exists a node $s'''$ such that $\prefix (s'',s''')$ and $\prefix (s''',s')$ and $s'''.state\in F$:
                      $s''$ is set as the backnode of $s'$, $s'$ is set as a leaf node, and the procedure jumps to the decision of whether the and branching succeeds in the $(j+1)^{st}$ direction, if
                      $j<k$, or finishes successfully the expansion of the tree beyond $s$ (Figure \ref{ftm}: lines \ref{ftm17}, \ref{ftm18}, \ref{ftm19} and \ref{ftm21})
        \end{enumerate}
(B) if there is no internal node $s''$ such that $s''.state=s'.state$ and $s''.PTPge=s'.PTPge$ then the and branching succeeds in the $j^{th}$ direction $\iff$
              the recursive call $ftm (s')$ succeeds (Figure \ref{ftm}: line \ref{ftm12})
%
%

Whenever the tree has been successfully expanded beyond node $s$ (Figure \ref{ftm}: line \ref{ftm21}), the procedure checks whether it has successfully built a whole tree,
by looking at whether $s$ is a rightmost leaf node, i.e., at whether $s.word\in\{d_k\}^*$ (Figure \ref{ftm}: line \ref{ftm22}):\\
(A) if it has, in other words, if $u\in\{d_k\}^*$, the boolean variable $gcsp\_checked$ tells the procedure whether consistency of the global CSP of the tree had been positively
           checked before. If it had been (Figure \ref{ftm}: line \ref{ftm222}), the procedure simply returns the information to the upper nodes, until it gets to the root, and then to the
           main program, which can then, thanks to the pointer variable $ftm\_root$, access and print the finite tree representation so constructed, of an accepting regular run of B. If
           it had not been, the global CSP is computed  (Figure \ref{ftm}: lines \ref{ftm23} and \ref{ftm232}) and its consistency checked:
    \begin{enumerate}
      \item In case of consistency, the procedure sets the boolean variable $gcsp\_checked$ to $true$, so that the upper nodes will not redo the work of checking the global CSP; and
               returns $true$ (Figure \ref{ftm}: line \ref{ftm24})
      \item In case of inconsistency, the current choice of $\delta (s.state)$ fails, the subtree at $s$ is cut from the tree being constructed (Figure \ref{ftm}: restore operation, line
               \ref{ftm242}), and the procedure jumps to the next choice of $\delta (s.state)$, if any, or backtracks with failure to expand the tree beyond $s$ (Figure \ref{ftm}: line
               \ref{ftm27})
    \end{enumerate}
(B) if the procedure has not finished yet the construction of a whole tree, it continues the attempt to do so (Figure \ref{ftm}: line \ref{ftm26}).

To finish, a word on the functions $globalcsp(s)$ and $variable(s,u)$ (Figure \ref{ftmmain}) is in order. The function $globalcsp(ftm\_root)$ is called when a whole finite tree model
has been successfully built  (Figure \ref{ftm}: line \ref{ftm23}), and computes, in a depth-first manner guided by the order $d_1<\ldots <d_k$, the global CSP of the tree, which is the union of the CSPs of the different
internal nodes. The CSP of internal node $s$, in turn, is the union $\bigcup_{P(u_1,\ldots ,u_r)\in s.Constraints}\{P(variable(s,u_1),\ldots ,variable(s,u_r)\}$. $variable(s,u_j)$, with
$u_j\in K^*N_{cF}$, is the variable (or parameter) targetted by $u_j$ at node $s$, and is written as the pair $<s',g>$ such that $u_j=d_{i_1}\ldots d_{i_\ell}g$ and $s'$ is the
$d_{i_1}\ldots d_{i_\ell}$-successor of $s$ in the tree. In the computation of $variable(s,u)$, the leaf nodes $s'$ are used as pointers to the internal nodes $s''$ verifying
$s''=\backnode (s')$ (see Figure \ref{ftmmain} for details). \cqfd
%

\section{Conclusion and future work}\label{conclusion}
We have thoroughly investigated B\"uchi automata augmented with spatial constraints. In particular, we
have provided a translation of an alternating into a nondeterministic, and an effective nondeterministic
doubly depth-first polynomial space algorithm for the emptiness problem of the latter.

A future work worth mentioning is whether
one can keep the same spatio-temporalisation of the other work and define a form of TBox cyclicity
stronger enough to subsume the semantics of the well-known mu-calculus, and make the latter benefit from the results of this work.
%
%
\newpage\noindent
\bibliographystyle{aaai}
\bibliography{biblio-c-maj}

\end{document}